\documentclass{aip-cp}

\def\to{\rightarrow}

\def\bi{\begin{itemize}}
\def\ei{\end{itemize}}

\def\ta{\tilde a}

\def\tg{\tilde g}

\def\tw{\widetilde W}
\def\tz{\widetilde Z}
\def\alt{\stackrel{<}{\sim}}
\def\agt{\stackrel{>}{\sim}}
\def\be{\begin{equation}}  
\def\ee{\end{equation}}  
\def\bea{\begin{eqnarray}}  
\def\eea{\end{eqnarray}}  

\usepackage[numbers]{natbib}
\usepackage{rotating}
\usepackage{graphicx}

\begin{document}

\title{(Mainly) axion dark matter\footnote{Plenary talk given at the Particle Physics and Cosmology 2015 (PPC2015) meeting, Deadwood, SD, June 29, 2015}}

\author[aff1]{Howard Baer\corref{cor1}}

\eaddress[url]{http://www.nhn.ou.edu/\~\ baer}

\affil[aff1]{Homer L. Dodge Dep't of Physics and Astronomy, University of Oklahoma, Norman, OK 73019, USA}
\corresp[cor1]{Corresponding author: baer@nhn.ou.edu}

\maketitle

\begin{abstract}
The strong CP problem of QCD is at heart a problem of naturalness: 
why is the $F\tilde{F}$ term highly suppressed in the QCD Lagrangian when it seems necessary 
to explain why there are three and not four light pions? 
The most elegant solution posits a spontaneously broken Peccei-Quinn (PQ) symmetry 
which requires the existence of the axion field $a$. 
The axion field settles to the minimum of its potential thus removing the offensive term 
but giving rise to the physical axion whose coherent oscillations can make up the cold dark matter. 
Only now are experiments such as ADMX beginning to explore QCD axion parameter space. 
Since a bonafide scalar particle-- the Higgs boson-- has been discovered, one might expect its mass
to reside at the axion scale $f_a\sim 10^{11}$ GeV. 
The Higgs mass is elegantly stabilized by supersymmetry: 
in this case the axion is accompanied by its axino and saxion superpartners. 
Requiring naturalness also in the electroweak sector implies higgsino-like WIMPs 
so then we expect mixed axion-WIMP dark matter. 
Ultimately we would expect detection of both an axion and a WIMP while signals for light higgsinos
may show up at LHC and must show up at ILC.
\end{abstract}

\section{Introduction: the strong CP problem and axions}

The story of the axion begins at the dawn of QCD, circa mid-1970s\cite{peccei}. With two light quarks, 
one expects QCD to manifest an approximate $U(2)_L\times U(2)_R$ chiral symmetry which can be recast 
as $U(2)_V\times U(2)_A$. The $U(2)_V$ symmetry gives rise to the well-known 
isospin and baryon number symmetries 
$SU(2)_I\times U(1)_B$ while the axial $U(2)_A$ is spontaneously broken. Since $U(2)_A$ is rank 4, 
we expect  four pseudo-Goldstone bosons-- the pions-- whilst we see only three with
$m_\eta\gg m_\pi$. Weinberg dubbed this conundrum the $U(1)_A$ problem and suggested that somehow
nature doesn't respect the $U(1)_A$ symmetry\cite{U1}. Shortly thereafter, 't Hooft discovered the
QCD $\theta$ vacuum and the effect of instantons\cite{tHooft}. 
Indeed, the ground state of QCD did not respect the
$U(1)_A$ symmetry and $m_\eta\gg m_\pi$ was explained. The price to pay was that the QCD Lagrangian
should contain a term of the form 
\be {\cal L}\ni\theta \frac{g_s^2}{32\pi^2}F_{A\mu\nu}\tilde{F}_A^{\mu\nu} .
\label{eq:Lqcd}
\ee
In addition, a complex quark mass matrix $M$ gives rise to a second contribution so that
$\theta\rightarrow\bar{\theta}\equiv \theta +Arg\ det M$. This term violates $CP$ symmetry
and leads to a measurable contribution to the neutron EDM. Measurements of the neutron EDM 
imply $\bar{\theta}\alt 10^{-10}$. Somehow the term must be very tiny. This is contrary to the old maxim
of quantum mechanics: {\it everything is allowed unless explicitly forbidden}.
How to suppress or get rid of the awkward Lagrangian term Eq. \ref{eq:Lqcd} 
constitutes the famous strong CP problem. 

While a variety of solutions to the strong CP problem have been proposed, 
one stands out for its simplicity and elegance: the Peccei-Quinn (PQ) solution\cite{pqww}. 
PQ proposed the existence of a new global $U(1)_{PQ}$ symmetry which is spontaneously broken 
at the PQ energy scale $f_a$. A new Goldstone field, the {\it axion}, arises in accord with Goldstone's 
theorem. The QCD Lagrangian now includes
\be
{\cal L}\ni\frac{1}{2}\partial^\mu a\partial_\mu a +\left(\frac{a}{f_a}+\bar{\theta}\right)
\frac{\alpha_s}{8\pi}F_{A\mu\nu}\tilde{F}_A^{\mu\nu}
\ee
where the latter term provides a potential for the axion field:
\be
V_{eff}\sim \left(1-\cos (\bar{\theta}+\frac{a}{f_a})\right) .
\ee
The axion field settles dynamically to its minimum at $\langle a\rangle=-f_a\bar{\theta}$
and the offending term goes to zero: the strong $CP$ problem is solved!
As a consequence of this solution, a physical axion particle should exist with mass\cite{bardtye}
$m_a^2=\langle\frac{\partial^2V_{eff}}{\partial a^2}\rangle$ where 
$m_a\sim 6\mu{\rm eV}\frac{10^{12}\ {\rm GeV}}{f_a}$. In the original proposal PQ suggested the scale
$f_a\sim m_{weak}$, a value which was soon ruled out by reactor and beam dump experiments. 
KSVZ\cite{ksvz} and DFSZ\cite{dfsz} proposed variant models with much higher values 
of $f_a\sim 10^{11}$ GeV, a value which suppressed anomalous decays leading to the so-called
{\it invisible} axion. Sikivie stripped off the invisibility cloak by suggesting axion detection in microwave
cavity experiments\cite{sikivie}. 

\section{Axion dark matter: relic density and detection}

\begin{figure}[h]
 \centerline{\includegraphics[width=350pt]{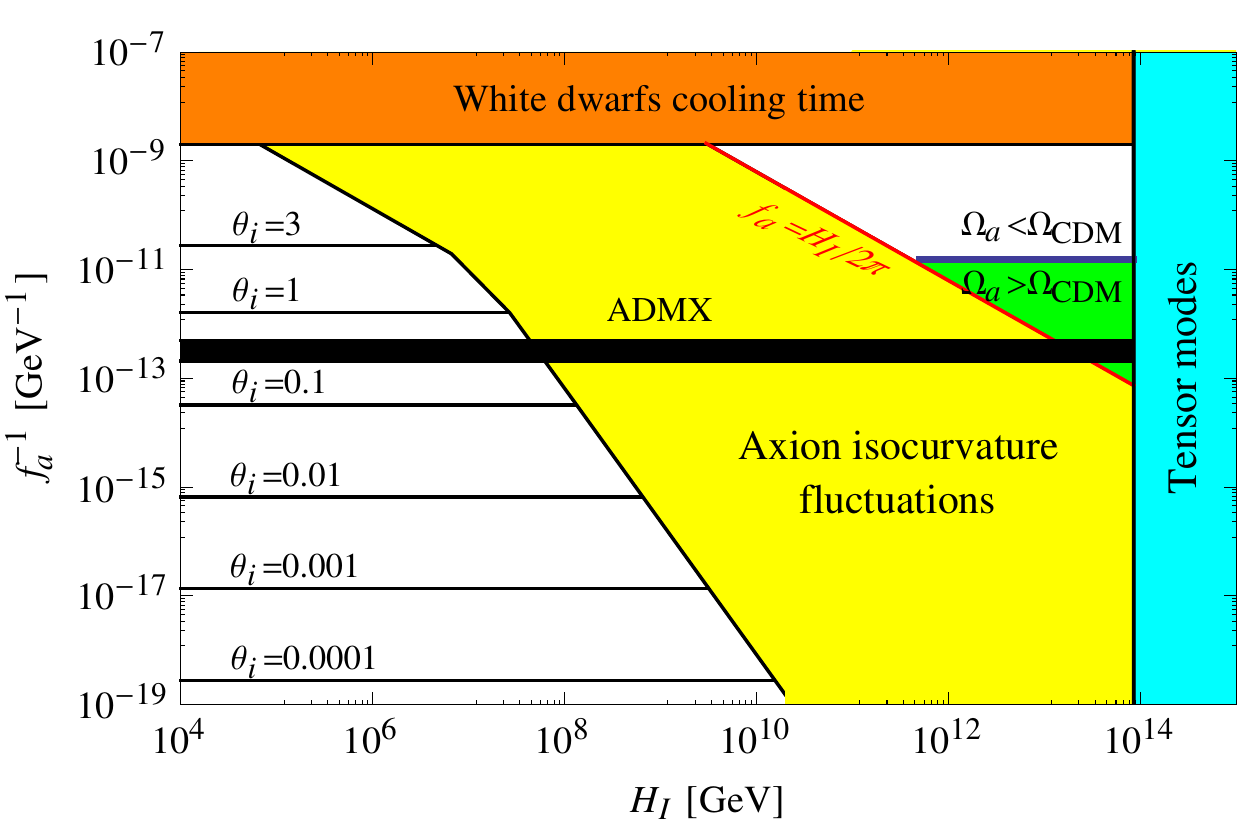}}
  \caption{Plot of $f_a^{-1}$ vs. $H_I$ parameter space for axion-only CDM. 
The un-shaded regions are all allowed as is the blue line where axions saturate 
the measured CDM relic density. The green region produces too much axion CDM.
The red line denoted $f_a=H_I/2\pi$ marks the boundary where PQ breaking takes place before (left) or after (right) the end of inflation.
The yellow region is excluded by axion isocurvature fluctuations. 
ADMX has explored the narrow region with 
$f_a\sim 2\times 10^{12}$ GeV. 
The left-side unshaded region is allowed for each $f_a$ value
by choosing an appropriate vacuum misalignment angle $\theta_i$. 
$\theta_i\sim 1$ seems most natural.
Figure from Ref. \cite{bckr}.
}
\label{fig:axion}
\end{figure}

In spite of the fact that axions are extremely weakly coupled-- 
their coupling strength is suppressed by $1/f_a$-- they still can play the role of cold dark matter.
The equation of motion for the axion field $\theta =a(x)/f_a$ in the early universe is given by
\be
\ddot{\theta}+3H(T)\dot{\theta}+\frac{1}{f_a^2}\frac{\partial V(\theta)}{\partial\theta}=0
\ee 
with $V(\theta )=m_a^2(T)f_a^2(1-\cos\theta )$. 
This is the equation of a damped harmonic oscillator.
At high temperatures, $m_a(T)\simeq 0$
and the solution is that $\theta\sim \theta_i$ a constant. The axion mass $m_a(T)$ turns on
for $T\alt 1$ GeV and the axion field begins coherent oscillations for which the equation of state is that
of cold dark matter. 
If PQ symmetry breaks after the end of inflation, then $\theta_i$ must be averaged over disparate domains
leading to a definite prediction of $\Omega_ah^2\sim 0.12$ shown by the blue line in Fig. \ref{fig:axion}.
If PQ symmetry breaks before the end of inflation, then the
axion relic density computed from this ``vacuum mis-alignment'' mechanism is given by\cite{axdm}
\be
\Omega_ah^2\sim \frac{1}{2}\left[\frac{6\times 10^{-6}{\rm eV}}{m_a}\right]^{7/6}\theta_i^2h^2
\ee
so that the measured relic density can be achieved for any $f_a$ value by an appropriate choice of $\theta_i$.
Values of $f_a\alt 10^9$ GeV are disallowed by stellar cooling bounds\cite{raffelt} 
while large values of Hubble constant at the end of inflation $H_I$ are disallowed by axion isocurvature
limits\cite{wilczek}: see Fig. \ref{fig:axion} for a panoramic view.

The region of Fig. \ref{fig:axion} around $f_a\sim 2\times 10^{12}$ GeV has been explored by the ADMX experiment. 
ADMX makes use of a microwave cavity which can be tuned over a range of frequencies.
If the frequency is just right, then resonance occurs and a bump should appear in the spectra.
ADMX has been upgraded with a new dilution refrigerator and SQUID electronics and hopes to explore
more broadly and deeply in $g_{a\gamma\gamma}$ vs. $m_a$ parameter space in the near future.

Expectations for an axion signal are model dependent. Two commonly used models include
KSVZ which introduced intermediate mass scale PQ charged heavy quark fields $Q$ or the DFSZ model
wherein two PQ charged Higgs doublets are required which couple to a PQ charged but SM singlet field $\chi$.
The ADMX experiment has reached sensitivity to axions in the KSVZ model but at present is short of the DFSZ
coupling strength.

\section{Supersymmetric axions: axinos and saxions}

While much of the previous discussion  is based on long-known results, the biggest recent development
in axion physics is.... the discovery of the Higgs boson with $m_h\simeq 125$ GeV\cite{atlas_h,cms_h}.
The issue here is that the Higgs boson exists and appears very much SM-like and as a fundamental scalar field.
If one then expands the SM to include axions via {\it e.g.} the KSVZ or DFSZ model, then one would 
expect the Higgs mass to blow up to $m_h\sim f_a$ scale. Of course, one can always fine-tune $m_h^2$ to 
$m_h^2/f_a^2\sim 10^{-20}$ but such tuning is generally symptomatic of some missing ingredient in the 
model.\footnote{Weinberg states: ``The appearance of fine-tuning in a scientific theory 
is like a cry of distress from nature, complaining that something needs to be better explained''\cite{sweinberg}.}

This electroweak fine-tuning crisis is elegantly solved via the introduction of
supersymmetry where the offending quadratic divergences to scalar field masses all neatly cancel.
Weak scale SUSY (SUSY with weak scale soft breaking) is supported experimentally by 1. the measured strengths 
of the gauge couplings which neatly unify within the MSSM\cite{gauge}, 
2. the large mass of the top quark which is needed for radiative EWSB\cite{rewsb} and 3. 
the measured value of $m_h$ which squarely sits in the predicted SUSY window $m_h\alt 135$ GeV\cite{mhiggs}.

Nonetheless, it has alarmed many theorists that SUSY matter states have not be found at LHC8 searches.
However, many other theorists expected a rather high value of soft breaking scale $m_{SUSY}$ since it offers a
decoupling solution to the SUSY flavor and CP problems and is in accord with the gravitino problem
for gravitino mass $m_{3/2}\agt 5$ TeV\cite{dine}. 
The enigma is the connection of the SUSY breaking scale to the weak scale:
$m_{weak}\sim 100$ GeV $\ll m_{SUSY}\agt 2$ TeV. This is the emerging Little Hierarchy problem.
It requires a scrutinization of naturalness measures.

While simple evaluations of large logs or Barbieri-Giudice naturalness $\Delta_{BG}$ pointed to 
high fine-tuning, the methodology used in these measures has been criticized\cite{comp,seige,mt,arno,xt}
in that they neglect {\it dependent contributions of opposite-sign} which lead to large cancellations.
A proper evaluation of fine-tuning allows for TeV-scale highly mixed 3rd generation squarks but does
require light higgsinos $\tw_1^\pm,\ \tz_{1,2}$ of mass $\mu\sim 100-300$ GeV, the lighter the better.
The lightest higgsino $\tz_1$ is the LSP but is thermally underproduced as WIMP dark matter
with typically $\Omega_{higgsino}^{TH}h^2\sim 0.007-0.01$.
The light higgsino spectrum is compressed with inter-higgsino mass gaps of $\sim 10-30$ GeV
so that higgsino decay products are soft and are buried under QCD backgrounds at LHC\cite{peisi}.
The only soft SUSY breaking term required to be near $\sim m_{weak}$ is $m_{H_u}^2$. This term can be driven 
radiatively  to small instead of large negative values at the weak scale 
in models with non-universal Higgs soft masses at the GUT scale. 
Such models are those with radiatively-driven naturalness or RNS\cite{rns}.

By requiring SUSY to solve the EW naturalness problem and PQ symmetry to solve the QCD naturalness
(strong CP) problem, then the axion becomes but one element of the axion superfield. It is now accompanied
by the spin-0 $R$-parity-even saxion $s$ and the spin-$1/2$ $R$-parity-odd axino $\ta$. 
In gravity-mediated SUSY breaking models, it is expected that $m_s\sim m_{\ta}\sim m_{3/2}$\cite{cl}.
In such a case, then dark matter is composed of two particles: the SUSY WIMP and the axion.

The presence of axions in weak scale SUSY models offers an important and 
elegant solution to the SUSY mu problem.
Since the $\mu$ term occurs in the superpotential (it is supersymmetric and not SUSY breaking) one expects naively
its value to be $\sim M_P$ (the Planck scale) while phenomenology/naturalness require $\mu\sim m_{weak}\sim 100$ GeV.
Kim and Nilles recognized\cite{kn} that the supersymmetrized DFSZ axion model offers a solution to the mu problem.
The mu term is first forbidden because it violates PQ symmetry but then it is regenerated via the
DFSZ coupling $W\ni\lambda_\mu X^2H_uH_d/M_P$ where the PQ field $X$ develops a vev $\langle X\rangle\sim f_a$
under PQ breaking. Then 
\be
\mu\sim \lambda_\mu f_a^2/M_P
\ee
while $m_{SUSY}\sim m_{hidden}^2/M_P$. The Little Hierarchy with $\mu\ll m_{SUSY} $ is just a reflection of
a mis-match between PQ breaking scale and hidden sector mass scale $f_a\ll m_{hidden}$. 
Such a mis-match can arise in models such as the MSY model of radiatively-driven PQ breaking\cite{msy}.
In the MSY model, SUSY breaking effects $\sim m_{3/2}$ drive one of the PQ fields to negative squared mass
causing PQ symmetry to break much as EW symmetry is broken radiatively due to the large top quark Yukawa coupling.
Minimizing the PQ scalar potential, then one typically finds $\mu\sim 100-300$ GeV is induced by 
$m_{3/2}\sim 2-20$ TeV\cite{radpq}: the mu problem is solved and the Little Hierarchy is explained! 
The PQ breaking scale $f_a$ sets the mass for the axion, the Higgs and the higgsinos!
In addition, Majorana masses are induced for right-hand neutrinos. 
While this behavior is exhibited for the MSY model, it seems typical of a much larger class of models.

\section{Signals at LHC and  ILC}

SUSY models with radiatively driven naturalness and with a supersymmetrized DFSZ axion solution
are natural in both the EW and QCD sectors, can accommodate $m_h\sim 125$ GeV with Tev-scale highly mixed stops
and can evade LHC8 searches. In contrast to the usual expectation that $m_{sparticles}\sim m_{weak}$, in this
case only $m_{higgsinos}\sim m_{weak}$-- the other sparticles can be much heavier, $\sim 2-20$ TeV.
Upper bounds on sparticle masses are computed in Ref. \cite{rns,upper}. 
For natural solutions with $\Delta_{EW}<10$, then $m_{\tg}<2$ TeV, within the reach of 
high-luminosity LHC13\cite{lhc}.
In this case, gluino cascade decay events should contain a characteristic dilepton mass edge\cite{lhc,baris} with
$m(\ell^+\ell^- )\alt 10-30$ GeV in accord with the inter-higgsino mass gap. Also, a unique same-sign diboson
signature sans hard jet activity should arise from wino pair production\cite{lhcltr,lhc} $pp\to \tw_2^\pm\tz_4$ where $\tw_2\to W\tz_{1,2}$ and
$\tz_4\to W^\pm\tw_1^\mp$. For $\Delta_{EW}<30$ (as defined in {\it e.g.} Ref. \cite{rns}), then $m_{\tg}$ can range up to 4 TeV and lie beyond LHC13 reach.
Monojet plus dilepton signals offer another LHC detection 
possibility\cite{kribs,lljet}.\footnote{For more on monojet signals, see also Ref's \cite{chan,mono,sasha}.}

The smoking gun signature of RNS SUSY will be direct higgsino pair production at an $e^+e^-$ collider
such as ILC\cite{ilc} which would operate with $\sqrt{s}>2\mu$. Built originally as a higgs factory, 
ILC will turn out to be a {\it higgsino factory}. SUSY can be discovered (or an LHC13 discovery can be confirmed)
and precision measurements can be made which test both the higgsino and gaugino sectors.

\section{Signals at axion and WIMP detectors}

\begin{figure}[h]
 \centerline{\includegraphics[width=350pt]{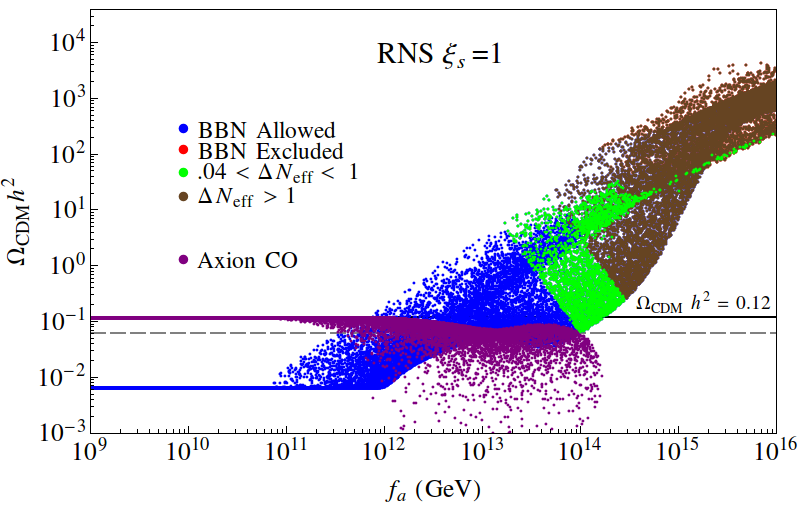}}
  \caption{ Plot of axion and higgsino-like WIMP relic abundance versus PQ scale $f_a$ in 
radiatively-driven natural SUSY with a DFSZ axion.
Figure from Ref. \cite{rnsdm}.
}
\label{fig:oh2}
\end{figure}

In SUSY with radiatively-driven naturalness, 
one expects a mixture of axion plus higgsino-like WIMP dark matter. 
The computation of the dark matter relic abundance requires the solution of eight coupled Boltzmann equations
which track the radiation, neutralino, axino,gravitino, saxion and axion (both CO- and TH-production) 
contributions. Axino production and decay can feed into and augment the neutralino abundance.
Saxion production and decay can feed the neutralino abundance or dilute it via decay to radiation; 
it can also inject dark radiation via $s\to aa$ 
decays.\footnote{See also Ref. \cite{rich}} 
If too many WIMPs are produced from axino or saxion decays, then they may re-annihilate at the
particle decay temperature\cite{az1}.
The calculation for the SUSY DFSZ model\cite{dfsz1} is given in Ref. \cite{dfsz2} and Fig. \ref{fig:oh2} for the DFSZ axion in natural SUSY. For low $f_a\sim 10^9-10^{11}$, 
higgsinos are underproduced and the relic abundance is axion-dominated\cite{bbc}. For higher $f_a$ values, the
axino and saxion decay later and increase the WIMP abundance. For too large $f_a$, then WIMPs become overproduced
and the model becomes excluded.

\begin{figure}[h]
 \centerline{\includegraphics[width=350pt]{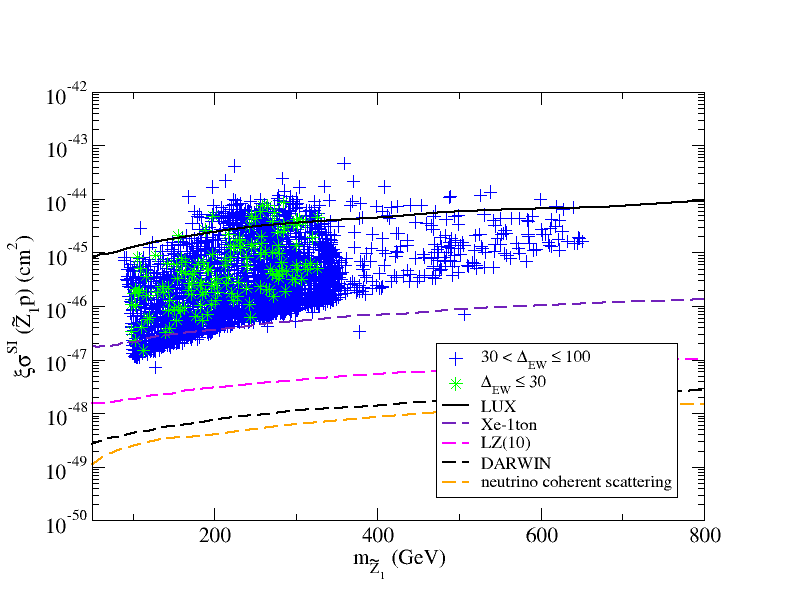}}
  \caption{ Plot of $\xi_s\sigma_{SI}(\tz_1 p)$ from radiatively-driven naturalness
where the local abundance is scaled down in accord with only a partial local abundance of DM coming from
WIMPs.
Figure from Ref. \cite{rnsdm}.
}
\label{fig:SI}
\end{figure}

What of WIMP detection? In models with mixed axion-WIMP dark matter, then WIMPs make up only part of the
relic abundance so the assumed local abundance of WIMPs must be scaled down by a factor
$\xi\equiv \Omega_{\tz_1}h^2/0.12$. The results of rescaled higgsino-like WIMP SI direct detection
rates are shown in Fig. \ref{fig:SI}. In spite of the rescaling, ton-scale noble liquid detectors should 
make a  complete exploration of the expected parameter space. While prospects for direct WIMP detection are excellent, 
rates for indirect detection must be rescaled by a factor of $\xi^2$ (for WIMP-WIMP annihilation in the
galactic halo) or by $\xi$ (for IceCube searches); these factors typically suppress 
detection rates to very low levels\cite{bbm,rnsdm}. 

\begin{figure}[h]
 \centerline{\includegraphics[width=350pt]{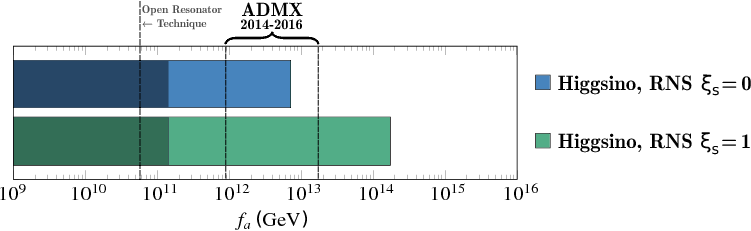}}
  \caption{ Plot of $f_a$ versus axion detection capabilities at ADMX and successors.
Figure from Ref. \cite{rnsdm}.
}
\label{fig:axdet}
\end{figure}

In addition to WIMP detection, we also expect detection of a DFSZ-like axion. 
The range of $f_a$ that can be explored by ADMX and successor experiments is shown in Fig. \ref{fig:axdet}. 

\section{Bullet point summary}

\begin{itemize}
\item Axion dark matter is a byproduct of the elegant PQWW/KSVZ/DFSZ
solution to the strong CP problem.
\item Axions need SUSY or else we would expect $m_h\sim f_a$. 
\item Electroweak naturalness requires the SUSY mu term $\mu\sim 100-300$ GeV, 
the lighter the better.
\item The SUSY DFSZ axion model allows for a solution to the SUSY mu problem.
\item The Little Hierarchy $\mu\ll m_{SUSY}$ may be a reflection of $f_a\ll m_{hidden}$.
\item Small $\mu\sim 100$ GeV can be generated from large $m_{3/2}$ in gravity-mediated SUSY breaking
models with radiative breaking of PQ symmetry.
\item LHC may discover natural SUSY but discovery is not guaranteed in SUSY with
radiatively-driven naturalness.
\item The ILC would make a guaranteed search for light higgsinos. Upon discovery, 
precision measurements of their properties are possible.
\item In such models, we expect mixed axion-higgsino-like WIMP dark matter.
\item Ultimately, we expect detection of both an axion and a WIMP.
\end{itemize}

\section{ACKNOWLEDGMENTS}
I thank CETUP and B. Szczerbinska for their kind hospitality.
I thank my collaborators Kyu Jung Bae, Vernon Barger, K. Y. Choi, 
J. E. Kim, Andre Lessa, Dan Mickelson, Azar Mustafayev, 
Maren Padeffke-Kirkland, Mike Savoy, Hasan Serce and Xerxes Tata.
This research was funded in part by the US Department of Energy Office of High Energy Physics.


%
\end{document}